\documentstyle[amssymb,preprint,aps]{revtex}
\tightenlines

\begin{document}
\title{Ferromagnetism in a (001), (110), and (111)-Oriented Ru Monolayer on 
Ag, Au,
and Cu-Substrates}
\author{L. M. Garc\'{\i}a-Cruz, A. V. Gaftoi, A. Rubio-Ponce}
\address{Departamento de Ciencias B\'{a}sicas, UAM-Azcapotzalco\\
Av. San Pablo 180, 02200 M\'{e}xico, D.F., MEXICO}
\author{A. E. Garc\'{\i}a}
\address{Escuela Superior de Ingenier\'{\i}a Qu\'{\i}mica e Industrias 
Extractivas,\\
IPN\\
Edificio 6, UP A. L\'{o}pez Mateos, Zacatenco, 07300 M\'{e}xico, D.F., 
MEXICO}
\author{R. Baquero}
\address{Departamento de F\'{\i}sica, CINVESTAV\\
A.P. 14-740, 07000 M\'{e}xico D.F. MEXICO}
\date{\today }
\maketitle

\begin{abstract}
We have studied the magnetic behavior of a 4$d$ transition metal Ru
monolayer (ML) on different substrates and orientations. In the ground
state, a Ru-ML is expected to be ferromagnetic on Ag(001) and Au(001) with a
magnetic moment $\left( \mu \right) $ of 1.73 Bohr magnetons $\left( \mu
_{B}\right) $ in both cases. On Cu(001), a Ru-ML is not magnetic. In this
paper, we study the magnetic behavior of a Ru-ML at other orientations, {\it 
i.e.}, (110) and (111). We found magnetism on Au (111), and Ag (111) ($\mu
\thicksim $1.3$\mu _{B}$ for both) but no magnetic activity on a Cu
substrate in any orientation. Further, we found that on Ag(110), a Ru-ML is
ferromagnetic with $\mu =1.3\mu _{B}$. On Au(110), a Ru-ML is not magnetic.
Since on the (001) and (111) orientations a Ru-ML has about the same
magnetic activity on Ag as on Au, we found surprising the behavior in the
(110) orientation. We analyze it in detail in the final part of the paper.
We found that there exists a metastable ferromagnetic state, in this case,
and that a Ru-ML becomes ferromagnetic under a small expansion of the
Au-lattice. This is not the case for Cu.
\end{abstract}

\section{INTRODUCTION}

In the 4$d$\ transition metal series, itinerant magnetism in three
dimensions (3D) was studied by Gunnarsson \cite{1} and Janak \cite{2} using
the qualitative Stoner criterium for ferromagnetism which is never satisfied
for any metal in the 4$d$ series \cite{3}. Electronic structure calculations
using state-of-the-art methods gave the interesting result that elements
might be forced to conserve their atomic magnetism, if synthesized at the
nanometer scale \cite{4}. In 2D, the coordination number is reduced and,
consequently, the $d$-band width changes and the local density of states
(LDOS) at the Fermi level, $\varepsilon _{F}$, increases \cite{5,6}.
Furthermore, the exchange integral, $J$, might in general, not be lower in
2D\ as compared to 3D\ \cite{7}. For these reasons, itinerant 2D\
ferromagnetism is not restricted to the elements that exhibit it in 3D\ 
\cite
{8,9,10}. In the past few years, several groups presented {\it ab- initio}
calculations on the ferromagnetism of 4$d$\ transition metal monolayers (ML)
on Ag(001) and Au(001) substrates. For the 4$d$ transition metals, ML
magnetism was predicted for Ru, Tc and Rh. For a Ru-ML a large magnetic
moment of 1.7 Bohr magnetons was predicted \cite{3,11}. No experimental
evidence of Ru-ML magnetism was given until Pfandzelter, Steierl and Rau
\cite{4} reported the first observation of 2D ferromagnetism on a Ru-ML
grown by epitaxy on a C(0001) substrate. Recently, Garc\'{i}a {\it et al.,}
\cite{12} studied Cu(001) as a substrate and obtained that a Ru-ML on
Cu(001) could become magnetic upon expansion of the Cu-lattice. They have
studied two factors to decide on whether the ML has or has not magnetic
activity. One is the lattice constant of the substrate which is adopted by
the ML grown on top of it. The other is the ML-substrate interaction of the
existing $d$-states. They have studied in detail the influence of expansion
and contraction of the ML-substrate distance. They have shown that when it
is contracted, and the interatomic distance of the Ru atoms on the
monolayer, consequently, enhanced, magnetism appears on the monolayer.

In this work, we study further the Ru-monolayer. We report the results
obtained for the magnetic moment on Ag , Au and Cu substrates in the (110)
and (111) orientations.

The rest of the paper is organized as follows. Section II is devoted to a
brief review of the main highlights of the theory. Section III is devoted to
our results. In the first part of it we report the magnetic moment resulting
for a Ru-ML from our calculations in the (110) and (111) orientations. We
have included the results in the (001) orientation from references \cite{3}
and \cite{12}, for completeness. Since the results in the (110) orientation
are somehow surprising, we include at the end of this section a detailed
discusion of them and study the effect of pressure and expansion in this
case. In the last section IV, we present our conclusions.

\section{\ THEORY}

We use the known surface Green function matching (SGFM) method \cite{n13} to
calculate the LDOS for the ML. To calculate the magnetic moment, we use both
the Hubbard \cite{13} method and a Stoner Hamiltonian \cite{14}. They both
lead to the same result for ferromagnetism.

\subsection{The Green's function}

The SGFM surface Green's function, $G_{s}$, is given by

\begin{equation}
G_{s}^{-1}(\varepsilon ,{\bf k})=\varepsilon I-H_{00}({\bf k})-H_{10}({\bf 
k}%
)T
\end{equation}
where $H_{00}$\ and $H_{01}$\ are the in-layer (surface) and interlayer
interaction Hamiltonians, respectively, in the customary description in
terms of principal layers. The $T$\ matrix is defined as $G_{10}=TG_{00}$. 
$%
G_{10}$\ is the propagator from the principal layer 0 to the first one. $%
G_{00}\equiv G_{s}$\ is the propagator within the surface principal layer. 
$%
I $\ is the unit matrix and $\varepsilon $\ is the energy. ${\bf k}$ is the
wave vector defined in the First Brillouin Zone (FBZ). A quickly converging
algorithm for the $T$\ matrix allows a very effective use of this and other
SGFM formulae\cite{2n13}.

We use Eq. (1) to calculate the LDOS for the (001)-, (111)- and (110)- free
surface. The SGFM surface Green's function, Eq. (1), has been used to
describe a substrate-ML system as well \cite{15,16,17,18}. Here, we rather
use the following formula for the Green's function, $G_{ML}$, of the
substrate-ML system \cite{12}.

\begin{equation}
G_{ML}^{-1}\left( \varepsilon ,{\bf k}\right) =G_{s(A)}^{-1}\left(
\varepsilon ,{\bf k}\right) +\left( \varepsilon I-H_{00(B)}({\bf k})\right)
-I_{A}H^{x}I_{B}-I_{B}H^{x}I_{A}
\end{equation}
In this supersupermatrix form, the expression is very transparent. The
supersupermatrix, $G_{ML}$, is labeled with the indices describing the two
media $A$\ and $B$. The upper diagonal part describes a semi-infinite medium
$A$\ with a surface (see Eq. 1); the lower diagonal part describes a
free-standing monolayer of $B$\ atoms. Both interact through the supermatrix
Hamiltonian $I_{M}H^{x}I_{N},(M,N=A,B)$. These matrices $H_{00(A)}$, $%
H_{10(A)}$, $H_{00(B)}$\ and $I_{M}H^{x}I_{N}$\ are readily written in the
two-center approximation within the Slater-Koster description \cite{19}.

The LDOS for the entire ML-substrate system is obtained from

\begin{equation}
N_{ML}\left( \varepsilon \right) =-\frac{1}{\pi }\int
%TCIMACRO{\func{Im}}%
%BeginExpansion
\mathop{\rm Im}%
%EndExpansion
\left[ {\rm Tr}G_{ML}\left( \varepsilon ,{\bf k}\right) \right] d{\bf k}
\end{equation}
We use the method by Cunningham to perform the numerical integration in the
2D FBZ \cite{20}.

\subsection{The magnetic moment}

The magnetic moment is calculated using first the Hubbard \cite{13} and then
the Stoner \cite{14} model. They both gave us the same result for a
ferromagnetic system. The magnetization, in units of Bohr magnetons, $\mu
_{B}$, is given by

\begin{equation}
\mu \left( \Delta \right) =\int\limits_{-\infty }^{\varepsilon _{F}}\left[
n_{d}^{+}\left( \varepsilon \right) -n_{d}^{-}\left( \varepsilon \right) %
\right] \ d\varepsilon =\int\limits_{\varepsilon _{F}-\frac{\Delta }{2}%
}^{\varepsilon _{F}+\frac{\Delta }{2}}\left[ n_{d}\left( \varepsilon \right) 
\right] \ d\varepsilon
\end{equation}
where $\Delta $\ is the magnetic band splitting, $n_{d}^{\pm }\left(
\varepsilon \right) $ indicates $n_{d}\left( \varepsilon \pm \frac{\Delta 
}{2%
}\right) $\ and, $n_{d}\left( \varepsilon \right) $\ is $d$-band
contribution to the paramagnetic density of states per spin, per eV, per
atom.

We conserve the total $d$-band electronic occupation, $N_{d}$, at each step,

\begin{equation}
N_{d}=\int\limits_{-\infty }^{\epsilon _{F}}\left[ n_{d}^{+}\left(
\varepsilon \right) +n_{d}^{-}\left( \varepsilon \right) \right] \
d\varepsilon
\end{equation}
so that charge transfer from the $p-$ or $s-$ sub-band is neglected.

The total energy, $E$, of the system, in this approximation, is calculated
from

\begin{equation}
E\left( \Delta \right) =\int\limits_{-\infty }^{\varepsilon _{F}}\left[
n^{+}\left( \varepsilon \right) +n^{-}\left( \varepsilon \right) \right] \
\varepsilon \ d\varepsilon +\frac{J\mu ^{2}}{4}
\end{equation}
where $n^{\pm }\left( \varepsilon \right) =n_{s}\left( \varepsilon \right)
+n_{p}\left( \varepsilon \right) +n_{d}^{\pm }\left( \varepsilon \right) $\
, where $n_{s}\left( \varepsilon \right) $\ and $n_{p}\left( \varepsilon
\right) $\ are the contributions to the LDOS from the $s$\ and $p$\ states,
respectively, and $J$ the Stoner parameter. In these equations the only
independent variable is $\Delta $. We get the magnetic moment from the value
of the magnetic band splitting, $\Delta _{0}$, that minimizes $E\left(
\Delta \right) $ in Eq.(6) with $\mu \left( \Delta \right) $ defined in Eq.
4.

\section{RESULTS}

\subsection{The density of states}

We have considered three different fcc substrates (Ag, Au, Cu) on top of
which a ML of the 4$d$-transition metal Ru, is grown. We consider the
substrate to be grown on three possible orientations, {\it i.e}., (001),
(110) and (111). A Ru-ML grown on Ag and Au in the (001) orientation has
been considered previously by Bl\"{u}gel \cite{3} and by Garc\'{i}a {\it et
al}., \cite{12} on a Cu substrate. We quote below their results for
completeness. The (110) and (111) cases are new.

First, we have calculated the total Ru-ML paramagnetic density of states for
each system and each orientation considered. In Table I, we give its value, 
$%
N(\varepsilon _{F})$, at the Fermi level. Magnetism is attributable to the
behavior of the $d$-electrons, and, therefore, their contribution to $%
N(\varepsilon _{F})$, should be, in principle, more significant. We quote
this value in Table I, as well. The trend is the same as expected and goes
as follows. The highest value is, in all the cases considered, the one for
the (001) orientation. For the (110) we get the smallest. Cu(001) is an
exception though, since $N(\varepsilon _{F})$ is highest in the (111)
orientation, instead. But since a Ru-ML is not magnetic on Cu in any
orientation (see below), we concentrate on the trend based on Ag and Au. For
any particular orientation, the system Ru/Ag presents the highest value of 
$%
N(\varepsilon _{F})$, Ru/Au follows, and Ru/Cu is the last.

\subsection{The magnetic moment}

As we stated above, the magnetic moment $\left( \mu \right) $\ is calculated
from the value of the magnetic splitting, $\Delta _{0}$, that minimizes $%
E(\Delta )$\ in Eq. (6). Our results appear in Table II. The magnetic
splitting, $\Delta _{0}$,\ obtained in this way correlates directly with $%
N(\varepsilon _{F})$ and with the corresponding $d-$contribution ($d-N\left(
\varepsilon _{F}\right) $) for any orientation, i.e., it is highest for
Ru/Ag, smaller for Ru/Au, and the smallest for Ru/Cu, where we get actually
zero. So, for a Ru-ML, $N(\varepsilon _{F})$, the $d-N(\varepsilon _{F})$,
and $\Delta _{0}$, have a common trend.

The Ru-ML spin discriminated total density of states (SLDOS) appears in
Figs. 1-3 for all the cases considered. On Table II, we give its value at
the $\varepsilon _{F}$ for the majority and minority spin.

>From the results presented in Figs. 1-3 and Eq. 4, we have calculated the
magnetic moment $\left( \mu \right) $\ at the Ru-ML. We give our results in
Table III. We use for the Ru-ML the Stoner parameter from Sigalas and
Papaconstantopoulus \cite{21}, $J=0.560$\ eV. From Table III, we see that $%
\mu $ follows the same trend as $N(\varepsilon _{F})$, $d-N(\varepsilon
_{F}) $, and $\Delta _{0}$. Indeed the highest moment is for Ru/Ag, Ru/Au
follows and Ru/Cu is last. Within each system, $\mu $ is highest in the
(001) orientation and lowest in the (110) one. For the (001)-orientation, we
find $1.73\mu _{B}$\ for Ru/Ag and Ru/Au, reproducing the results by
Bl\"{u}gel \cite{3}. For Ru/Cu we find zero \cite{12}. For (111), the
magnetic moment on Ru/Ag and Ru/Au are about the same, roughly $1.3\mu _{B}$
(See Table III). For Ru/Cu we find zero. These results are to be expected on
the grounds of the lattice constant value for these isoelectronic noble
transition metal substrates. The lattice constant for Ag and Au is
approximately equal, {\it i.e.}, 4.8\AA , while Cu has a lattice constant
equal to 3.61 \AA . Therefore Cu, in general, in any orientations shrinks
the space between the Ru-atoms, at the ML atomic layer, an effect that is
against the magnetic activity as it is discussed in Ref. \cite{12}. The
Ru-atoms on Cu have an interatomic distance which is shorter than the one on
the corresponding Ru- surface (the Ru-lattice constant is 3.81$%
%TCIMACRO{\unit{\AA }}%
%BeginExpansion
\mathop{\rm \AA }%
%EndExpansion
$). The Ru-surface is not magnetic in any orientation, according to our
calculation.

The trend that within each system $\mu $ is highest on the (001) orientation
and lowest in the (110) one, seems, at first sight, to be easily related to
a geometric property. But this is not actually the case. Let us try, for
example, to find a simple relation that agrees with the result that $\mu $
is smaller in the (111)-orientation than in the (001)one. In an fcc-lattice
the first-nearest neighbors (FNN) distance is $\frac{a}{\sqrt{2}}$, with 
$a$%
\ the lattice constant. When a monolayer is grown on top of a substrate, the
number of FNN that an atom on the ML has, varies with the crystallographic
orientation. Some of these neighbors are on the ML-atomic layer and others
belong to the substrate. Table IV summarizes these observation. The numbers
are presented following the trend that we got for the magnetic moment value.
We see from this Table that neither the total number of FNN, nor the number
of them on the ML-atomic layer or in the substrate correlate with the
magnetic moment. The last column line in Table IV is the distance between
the ML-atomic plane and the first substrate atomic plane. This parameter
does not correlate with $\mu $, as well. It seems that there is no simple
geometric parameter to correlate with the trend of the value that $\mu $\
takes in the different crystallographic directions.

As a first conclusion, we can say that the magnitude of the magnetic moment
on a Ru-ML grown on a noble metal depends strongly on the lattice constant.
It is about the same on Ag and Au and it is zero on Cu in any orientation.
Within each system the magnitude of $\mu $ is higher in the (001)
orientation and smallest on the (110). The trend of $\mu $ correlates with
the ML-total-paramagnetic density of states at the Fermi-level $N\left(
\varepsilon _{F}\right) $, with the corresponding $d-$band contribution to
it, and with the corresponding magnetic band splitting, $\Delta _{0}$.

There is an exception to this picture worth looking at. For a Ru-ML/Au(110),
$\mu =0$, while $\mu =1.3\mu _{B}$ for a Ru-ML/Ag(110). We devote next
sub-section to examine this point in more detail.

\subsection{\ The (110) orientation}

The null value for the Ru-ML/Au system in the (110)-orientation deserves
some attention. On general grounds the behavior of a Ru-ML on a Au or a Ag
substrate should not present a big difference in what the magnetic activity
is concerned. Actually a Ru-ML has essentially the same value of $\mu $ on
Au and on Ag in the (001)- and (111)-orientation. For this reason, the big
difference that we find in the (110) orientation for Ru/Ag and Ru/Au is
striking. The fact that the Ru-ML shows no magnetic activity on a
Cu-substrate in any orientation whatsoever is not, since, in this case, the
Ru atoms are brought together too closely by the substrate potential for any
magnetic activity to appear. But this argument does not hold true in the
case for Ru-ML/Au.

To further explore the difference between a Ru-ML/Au and a Ru-ML/Ag in the
(110) orientation, we have produce Fig. 4. It shows the difference curve
between the $d-$band contribution to the Ru-ML total paramagnetic density of
states, $N\left( \varepsilon _{F}\right) $, comparing a Ru-ML/Cu(110) to a
Ru-ML/Ag(110) (upper curve) and a Ru-ML/Au(110) to a Ru-ML/Ag(110) (lower
curve). In both cases, it is evident that the number of $d-$electrons, in
the vicinity of the $\varepsilon _{F}$, is much higher for the Ru-ML/Ag (the
difference curve is strongly negative) which is the only one that shows
magnetic activity in this direction. So the effect is due to the presence of
$d$-electrons much more nearby $\varepsilon _{F}$. Notice that the Stoner
criterium for ferromagnetism is fulfilled by Ru-ML/Ag(110) but it is not
neither by the Ru-ML/Au(110) nor by the Ru/Cu(110). Actually Ru on a Cu
substrate does not fulfill the Stoner criterium in any orientation
whatsoever. (See Table I and recall that, $J=0.560%
%TCIMACRO{\unit{eV}}%
%BeginExpansion
\mathop{\rm eV}%
%EndExpansion
$)

\subsection{The Effect of pressure and expansion in the (110)-orientation}

As a last point in this paper, we study the effect on the ML magnetism of
pressure and expansion of the Ru/Au (110) system. First, we present in Fig.
5 the total paramagnetic LDOS for the three systems considered. On the
top-graph the Ag(110) surface LDOS is compared to the Ru(110) surface LDOS
and to the Ru-ML/Ag (110) LDOS. The middle and lower graphs deal in the same
way with the Ru/Au(110) and Ru/Cu(110) cases. The case of interest here is
the Ru/Au(110). The other two are presented for completeness. Let us refer
to the middle graph (Ru/Au(110)). At $\varepsilon _{F}$ (the origin in the
graph), the Ru-ML/Au(110) LDOS is well below that the corresponding value
for Ru-ML/Ag(110) . Notice, nevertheless, that the peak, just below $%
\varepsilon _{F}$, in the Ru-ML/Au(110) LDOS is  almost as high as the one
on the corresponding curve for the Ru-ML/Ag(110) LDOS at $\varepsilon _{F}$.
So the question arises: can pressure or expansion bring this peak to the $%
\varepsilon _{F}$ and turn on the magnetic activity of the Ru-ML on a Au
(110) substrate?

The answer to this question appears in Fig. 6 and Table V. Pressure does not
turn on the magnetic activity but expansion does (enhancement of the Au
lattice constant). In this figure we present the total energy (Eq. 6) as a
function of the magnetic band splitting $\Delta _{0}$. Notice first that,
according to our results, a metaestable magnetic state does exist for a
Ru-ML/Au(110) at zero expansion (see the minimum in Fig. 6(a)). The
corresponding magnetic moment is $\mu =0.69\mu _{B}$\ and the magnetic
splitting is $\Delta _{0}=0.386$\ eV. This state is somehow premonitory of
the magnetic activity under expansion. In the rest of Fig. 6, we present the
magnetic activity of Ru/Au(110) under hydrostatic expansion up to 4\% of the
lattice constant of the substrate. We give, for completeness, a broader
account of our results in Table V. This kind of expansion might be possible
by applying pressure perpendicular to the sample since according to the
strain tensor this will enhance the distance between the Ru-atoms on the ML
atomic layer.

\section{\ CONCLUSIONS \newline
}

We have studied the magnetic activity of a Ru-ML grown on three different
substrates, {\it i.e}., Ag, Au and Cu in three crystallographic
orientations, (001), (111), and (011). In the (001) orientation, we
reproduced the results already obtained by Bl\"{u}gel \cite{3} and by
Garc\'{i}a {\it et al}., \cite{12}. The results for the other two directions
are new. We get a higher magnetic moment, $\mu $, for Ru/Ag, next Ru/Au and
last Ru/Cu. For each system, $\mu $ is highest on the (001) orientation and
lowest on the (110)-one. The Ru/Cu system has no-magnetic activity in any
orientation. For Ru/Ag, $\mu =1.73\mu _{B}$\ in the (001), $\mu =1.3\mu 
_{B}$%
\ in the (111), and $\mu =1.3\mu _{B}$\ in the (110). For the Ru/Au, $\mu
=1.73\mu _{B}$\ in the (001), $\mu =1.29\mu _{B}$\ in the (111), and $\mu 
=0$%
\ in the (110). See Table III and Figs. 1-3. The non existence of magnetic
activity for the Ru/Au (110) contrasts strongly with the Ru/Ag (110)
magnetic moment value of $\mu =1.3\mu _{B}$. To better understand this fact,
we have explored the behavior of the Ru/Au (110) system and found that a
metastable state does exist at zero pressure and that it becomes magnetic
under expansion (enhancement of the in-layer lattice constant). See Figs.
4-6 and Table V.

\section{ACKNOWLEDGMENTS}

We appreciate the interest of Prof. Carl Rau (Rice University) in our work
and the interesting comments from Prof. Leonard Kleinmann (University of
Texas ).

\newpage

Table I

\bigskip

\begin{tabular}{cc}
\hline\hline
Orientation & $N(\varepsilon _{F})$[states/spin/eV/atom] \\ \hline
\begin{tabular}{l}
\\
\\ \hline
001 \\
111 \\
110
\end{tabular}
&
\begin{tabular}{ccc}
Ag & Au & Cu \\
\begin{tabular}{cc}
total & $d$ \\ \hline
2.7042 & 2.6601 \\
2.6029 & 2.5899 \\
2.3390 & 2.3152
\end{tabular}
&
\begin{tabular}{cc}
total & $d$ \\ \hline
2.5824 & 2.5440 \\
2.5128 & 2.4948 \\
1.7469 & 1.7303
\end{tabular}
&
\begin{tabular}{cc}
total & $d$ \\ \hline
1.5132 & 1.4516 \\
1.5521 & 1.5235 \\
1.1665 & 1.1231
\end{tabular}
\end{tabular}
\\ \hline\hline
\end{tabular}

\bigskip

\bigskip

\bigskip

Table II

\bigskip

\begin{tabular}{cc}
\hline\hline
Orientation & $N(\varepsilon _{F})$[states/spin/eV/atom] \\ \hline
\begin{tabular}{c}
\\
\begin{tabular}{c}
\\ \hline
001 \\
111 \\
110
\end{tabular}
\end{tabular}
&
\begin{tabular}{ccc}
Ag & Au & Cu \\
\begin{tabular}{ccc}
maj. & min. & $\Delta _{0}\left( \text{eV}\right) $ \\ \hline
0.509 & 1.899 & 0.969 \\
0.701 & 1.744 & 0.725 \\
1.082 & 3.036 & 0.725
\end{tabular}
&
\begin{tabular}{ccc}
maj. & min. & $\Delta _{0}\left( \text{eV}\right) $ \\ \hline
0.531 & 1.723 & 0.972 \\
0.644 & 1.687 & 0.718 \\
1.715 & 1.715 & 0.000
\end{tabular}
&
\begin{tabular}{ccc}
maj. & min. & $\Delta _{0}\left( \text{eV}\right) $ \\ \hline
1.513 & 1.513 & 0.000 \\
1.547 & 1.547 & 0.000 \\
1.130 & 1.130 & 0.000
\end{tabular}
\end{tabular}
\\ \hline\hline
\end{tabular}

\bigskip

\bigskip

\bigskip

Table III

\bigskip

\begin{tabular}{cc}
\hline\hline
Orientation & The Magnetic Moment \\
substrate & $\left[ \mu _{B}\right] $ \\ \hline
\begin{tabular}{c}
\\
001 \\
111 \\
110
\end{tabular}
&
\begin{tabular}{ccccc}
Ag & \hspace{0.3in} & Au & \hspace{0.3in} & Cu \\ \hline
1.73 &  & 1.73 &  & 0.00 \\
1.30 &  & 1.29 &  & 0.00 \\
1.30 &  & 0.00 &  & 0.00
\end{tabular}
\\ \hline\hline
\end{tabular}

\newpage

Table IV

\bigskip

\begin{tabular}{cccc}
\hline\hline
Orientation & ML & Substrate & ML-first plane distance \\ \hline
001 & 4 & 4 & $\frac{a\ }{2}$ \\
111 & 6 & 3 & $\frac{a\ }{\sqrt{3}}$ \\
110 & 2 & 5 & $\frac{a\ }{2\sqrt{2}}$ \\ \hline\hline
\end{tabular}

\bigskip

\bigskip

\bigskip

Table V.

\bigskip

\begin{tabular}{cc}
\hline\hline
\% Expansion & Magnetic moment $\left[ \mu _{B}\right] $ \\ \hline
0 & 0.00 (0.69 metaestable state) \\
1 & 0.97 \\
2 & 1.04 \\
3 & 1.11 \\
4 & 1.18 \\
5 & 1.19 \\
6 & 1.32 \\
7 & 1.38 \\
8 & 1.43 \\
9 & 1.47 \\ \hline\hline
\end{tabular}

\newpage

\begin{center}
FIGURES CAPTIONS
\end{center}

\bigskip

\noindent Fig. 1 Our results for the Ru-ML/Ag in the three different
orientations. We present the Ru-ML spin discriminated total density of
states (SLDOS). In this case $\Delta _{0}\neq 0$ for all the three
orientations and, consequently, $\mu $ is also different from zero. See text
and Tables II and III.

\bigskip

\noindent Fig. 2 A Ru-ML/Au (See caption of Fig.1). Notice that, in this
case $\Delta _{0}=0$ for the substrate grown in the (110) orientation and
therefore $\mu =0$. Compare with the previous case. (See text).

\noindent Fig. 3 A Ru-ML/Cu. (See caption of Fig.1). $\Delta _{0}=0$ ($\mu
=0 $) for all the orientations in this case.

\bigskip

\noindent Fig. 4 Here we show the difference-curve between the $d-$band
contribution to the Ru-ML paramagnetic density of states for the Ru/Cu and
Ru/Ag (upper curve) and for the Ru/Au and Ru/Ag (lower curve) in the (110)
orientation. See text.

\bigskip

\noindent Fig. 5 We present three sets of graphs, one for each of the three
cases considered here [Ru-ML/Ag(110), Ru-ML/Au(110), and Ru-ML/Cu(110)]. For
each case we compare three curves that represent the total paramagnetic
local density of states (LDOS) for

i) The Ru-ML.

ii) The (110)-oriented surface of the substrate.

iii)\bigskip\ The (110)-oriented Ru surface.

\noindent Fig. 6 The total energy, $E\left( \Delta \right) $, (Eq. 6) as a
function of the magnetic band splitting, $\Delta $, for a Ru-ML/Au(110)
under expansion of the Au lattice constant. See text for details.

\newpage

\begin{center}
TABLES CAPTIONS
\end{center}

\bigskip

\noindent Table I. We present here the Ru-ML total paramagnetic density of
states at the $\varepsilon _{F}$, $N\left( \varepsilon _{F}\right) $, in the
orientations considered. The largest is for Ag, then for Au and the smallest
is for the Cu substrate. Also, for a given substrate, the highest $N\left(
\varepsilon _{F}\right) $, is in the (001) -orientation and the smallest in
the (110)-one. In the second column, we also quote the corresponding
contribution of the $d$-band. The trends are the same. Notice that for the
Cu-substrate, the trend is not followed in the (111)-orientation. Ru/Cu is
not magnetically active in any orientation.

\bigskip

\noindent Table II. In the first column, we present the corresponding
direction of growth of the substrate. Next, we quote our values for the
Ru-ML spin discrimated total density of states (SLDOS) at the Fermi level
for majority and minority spin. The next column gives the corresponding
magnetic band splitting, $\Delta _{0}$. See text.

\bigskip

\noindent Table III. We give the magnetic moment (in units of Bohr
magnetons, $\mu _{B} $), on the Ru-ML in different orientations for the
three substrates considered.

\bigskip

\noindent Table IV. We give here the number of first nearest neighbors (FNN)
of a Ru-ML atom that lie on the ML and in the substrate. The last column
gives the ML- first plane distance in each case. None of these parameters
follows the magnetic moment trend.

\bigskip

\noindent Table V. We present here the magnetic moment for a Ru-ML/Au(110)
under expansion of the Au lattice constant. See text for details.


\begin{references}
\bibitem{1}  O. Gunnarson, J. Phys. F {\bf 6}, 587 (1976).

\bibitem{2}  J. F. Janak, Phys. Rev. B {\bf 16}, 225 (1977).

\bibitem{3}  S. Bl\"{u}gel, Phys. Rev. Lett. {\bf 68}, 851 (1992).

\bibitem{4}  R. Pfandzelter, G. Steierl, and C. Rau, Phys. Rev. Lett. {\bf 
74%
}, 3467 (1995).

\bibitem{5}  L. M. Falicov, R. H. Victora, and J. Tersoff, in The Structure
of Surfaces, edited by M. A. van Hove and S. Y. Tong, Vol. 2 of Springer
Series in Surface Science (Springer, Berlin. 1985).

\bibitem{6}  G. Allan, Phys. Rev. B {\bf 19}, 4774 (1979).

\bibitem{7}  S. Onishi, C. L. Fu, and A. J. Freeman, J. Magn. Magn. Mater.
{\bf 50}, 161 (1985).

\bibitem{8}  S. Bl\"{u}gel, Europhys. Lett. {\bf 7}, 743 (1988); S.
Bl\"{u}gel, B. Drittler, R. Zeller, and P. H. Dederichs, Appl. Phys. A:
Solids Surf. {\bf 49}, 547 (1988).

\bibitem{9}  L. M. Falicov {\it et al}., J. Mater. Res. {\bf 5}, 1299 
(1999).

\bibitem{10}  C. L. Fu, A. J. Freeman, and T. Oguchi, Phys. Rev. Lett. {\bf 
%
54}, 2700 (1985).

\bibitem{11}  R. Wu and A. J. Freeman, Phys. Rev. B {\bf 45}, 7222 (1992).

\bibitem{12}  A. E. Garc\'{i}a, V. Gonz\'{a}lez-Robles, and R. Baquero,
Phys. Rev. B {\bf 59}, 9392 (1999).

\bibitem{n13}  F. Garc\'{i}a-Moliner and V. R. Velasco, Prog. Surf. Sci.
{\bf 21}, 93 (1986); R. Baquero, V. R. Velasco and F. Garc\'{i}a-Moliner,
Phys. Scripta {\bf 38}, 742 (1988); R. Baquero and A. Noguera, Rev. Mex. de
F\'{i}s. {\bf 35}, 638 (1989); C. Quintanar, R. Baquero, V. Velasco and F.
Garc\'{i}a-Moliner, Rev. Mex. de F\'{i}s. {\bf 35}, 742 (1988).

\bibitem{13}  J. Hubbard Proc. Roy. Soc., {\bf A276}, 238 (1963).

\bibitem{14}  S. V. Vonsovski, Magnetism Vol. 2, Wiley, New York, (1974).

\bibitem{2n13}  M. P. L\'{o}pez-Sancho, J. M. L\'{o}pez-Sancho and J. Rubio,
J. Phys. C: Met. Phys. {\bf 14}, 1205(1984); J. Phys. C: Met. Phys. {\bf 
15}%
, 855(1985)

\bibitem{15}  J. Mart\'{\i}n Y\'{a}\~{n}ez and R. Baquero, Rev. Mex.
F\'{\i}s. {\bf 40}, 287 (1994).

\bibitem{16}  R. de Coss and R. Baquero, Rev. Mex. F\'{\i}s. {\bf 41}, 875
(1995); R. de Coss, Phys. Rev. B {\bf 52}, 4768 (1995).

\bibitem{17}  R. de Coss, Ph.D. thesis, CINVESTAV, (1996).

\bibitem{18}  V. M. Gonz\'{a}lez Robles, Ph.D. thesis, CINVESTAV, (1997); A.
E. Garc\'{i}a, Ph.D. thesis, CINVESTAV, (1997).

\bibitem{19}  J. C. Slater and G. F. Koster, Phys. Rev. {\bf 94}, 1498
(1954).

\bibitem{20}  S. L. Cunningham, Phys. Rev. B {\bf 10}, 4988 (1974).

\bibitem{21}  M. M. Sigalas and D. A. Papaconstantopoulus, Phys. Rev. B {\bf 
%
50}, 7255 (1994).
\end{references}
\end{document}